\documentclass[a4paper]{ESASPCS13Style}
\usepackage{epsfig}

\begin{document}

\title{The shock-patterned solar chromosphere in the light of ALMA}

\author{S. Wedemeyer-B\"ohm\inst{1} \and
        H.-G. Ludwig\inst{2} \and
        M. Steffen\inst{3} \and  
        B. Freytag\inst{4} \and
        H. Holweger\inst{5}} \institute{
        Kiepenheuer-Institut f\"ur Sonnenphysik, Freiburg, Germany \and
        Lund Observatory, Lund, Sweden \and 
        Astrophysikal. Institut Potsdam, Potsdam, Germany \and 
        GRAAL, Universit\'e de Montpellier II, Montpellier, France \and 
        Inst.f.Theo.Phys.u.Astrophysik, Universit\"at Kiel, Germany}
      
\maketitle 

\begin{abstract}
Recent three-dimensional radiation hydrodynamic simulations by 
\cite{wedemeyer04} suggest that the solar chromosphere is highly structured in 
space and time on scales of only 1000~km and 20-25~sec, resp.. The resulting 
pattern consists of a network of hot gas and enclosed cool regions which are due 
to the propagation and interaction of shock fronts. In contrast to many other 
diagnostics, the radio continuum at millimeter wavelengths is formed in LTE, and 
provides a rather direct measure of the thermal structure. It thus facilitates 
the comparison between numerical model and observation. While the involved time 
and length scales are not accessible with todays equipment for that wavelength 
range, the next generation of instruments, such as the Atacama Large Millimeter 
Array (ALMA), will provide a big step towards the required resolution. Here we 
present results of radiative transfer calculations at mm and sub-mm wavelengths 
with emphasis on spatial and temporal resolution which are crucial for the 
ongoing discussion about the chromospheric temperature structure.

\end{abstract}

%================================================================================
\section{Introduction}
\label{sec:intro}

The structure of the solar chromosphere, and in particular of internetwork 
regions, is still an issue of intensive scientific debate. On the one hand UV 
diagnostics imply a hot layer with only small temperature fluctuations, as 
expressed in static semi-empirical 1-D models like, e.g., FAL~A, VAL~C 
(\cite{fal93}; \cite{val81}).
These models feature a temperature minimum and above a smooth increase towards 
high coronal values. 
In contrast the observed amount of carbon monoxide requires cool gas to be 
formed. Appropriate models (see, e.g., \cite{ayres02}) thus show very low 
temperatures at heights where the aforementioned class of models predicts much 
higher values. It should be noted that recent works 
(\cite{asensio03}; \cite{uitenbroek00a}; \cite{wedemeyer04})
suggest that CO is mostly located in the photosphere and low chromosphere, 
and its presence thus does not necessarily indicates low temperatures in the 
higher layers.  

The different models are very elaborate and explain the observations, which they 
were constructed for, very well. However, both type of models cannot provide a 
complete description of the solar chromosphere since they fail to explain the 
entirety of all available observations at the same time. 

A solution might be a very inhomogeneous and dynamic chromosphere 
which does exhibit cool and hot temperatures next to each other. Examples are 
the recent 3D radiation hydrodynamic simulation by Wedemeyer et al. (2004, 
hereafter Paper~I) and the pioneering 1-D work done by Carlsson \& Stein 
(e.g., 1995). Both show a very dynamic picture produced by propagating shock 
waves. Moreover, the mentioned 3D model reveals a network-like pattern of hot 
and cool gas on spatial scales which are of the same size as the underlying 
granulation pattern. 
 
In order to confirm the structure to be inhomogeneous and dynamic as implied by 
these models, detailed comparisons with observations are mandatory. But still 
such direct comparisons are hampered by various difficulties. One major problem, 
e.g. in the UV, is the breakdown of the LTE (local thermodynamic equilibrium) 
assumption at chromospheric heights, causing the translation of temperature into 
intensity to be a rather involved process. Hence, the production of synthetic 
intensity images from numerical models, which can be compared with observations, 
requires detailed multi-dimensional non-LTE radiative transfer calculations. 
Such calculations are hardly feasible at present, since they require large 
computational resources and efforts. 

The situation is physically simpler for the radio continuum at millimeter and  
sub-millimeter wavelengths which can be exploited as a linear measure 
of gas temperature, owing to the validity of LTE in this wavelength range.
The necessary radiative transfer calculations can be done 
comparatively easily. Hence, (sub-)millimeter wavelengths offer a convenient way 
to compare numerical models and observations as demonstrated in a recent work by 
\cite*{loukitcheva04}. 
Unfortunately, no appropriate observations are available so far since the spatial 
resolution of the existing instruments for this wavelength domain is too low,  
rendering the desired granular scales unaccessible for now. 

The situation will substantially improve in the future, when the Atacama Large 
Millimeter Array  (ALMA) will commence operation ($\sim 2011$). This instrument 
will provide high spatial and temporal resolution, allowing to observe the 
small-scale structure of the solar chromosphere.

Here we present images of emergent continuum intensity at millimeter wavelengths
which are calculated from the recent three-dimensional radiation hydrodynamics 
simulation by \cite*{wedemeyer04}. Since the simulation does not include 
magnetic fields, the presented analysis refers to non-magnetic internetwork 
regions only. Our aim is to predict how those areas would be seen by instruments 
like ALMA. 

A detailed paper with supplementary calculations is in preparation. 

%-------------------------------------------------------------------------------- 
\begin{figure}[t] 
\centering 
  \resizebox{6.14cm}{!}{\includegraphics{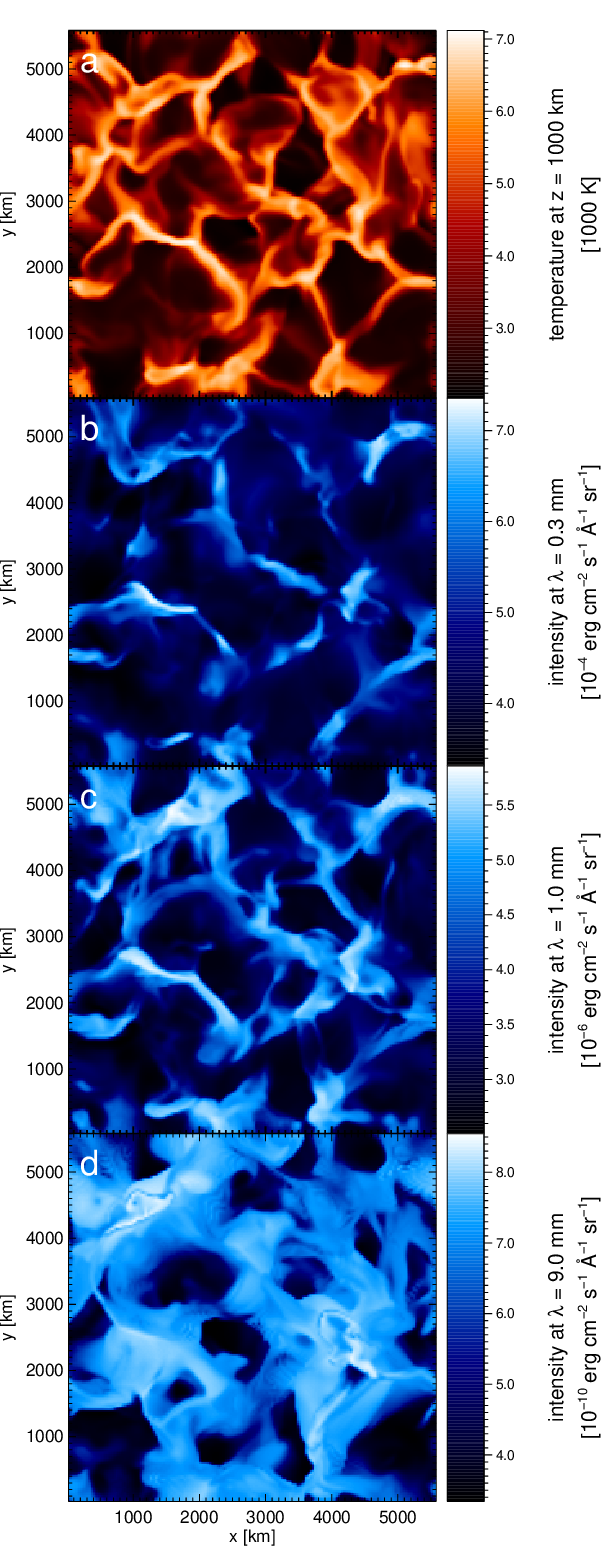}}
  \caption{Exemplary horizontal maps, all calculated from the same 
    simulation snapshot:  
    \textbf{a)} gas temperature at a geometrical height of $z = 1000$~km, and 
    \textbf{b-d)} emergent continuum intensity at wavelengths of $0.3$~mm, 
    $1$~mm, and $9$~mm, respectively. The legend next to each panel indicates 
    data range and colour-coding. All images are calculated for disk-center.
  } 
  \label{fig.intimages} 
\end{figure} 
%-------------------------------------------------------------------------------- 

%================================================================================
\section{Atacama Large Millimeter Array}
\label{sec:alma}

The Atacama Large Millimeter Array (ALMA) will consist of 64~telescopes with 
diameters of 12~m each, setting up 2016 baselines. The full array will start 
operation in 2011 in the Chilean Andes. There will be ten frequency bands 
which cover a total range from 31.3~GHz ($\lambda = 9.58$~mm) to 950~GHz 
($\lambda = 0.32$~mm). According to \cite*{bastian02}, ALMA will provide an 
angular resolution of $0.\hspace*{-0.7mm}\arcsec015$ to 
$1.\hspace*{-0.7mm}\arcsec4$, depending on antenna configuration. This 
corresponds to only  $\sim$~10~km to $\sim$~1000~km on the Sun. 

However, ALMA will not sample both these small spatial scales and the
larger scales needed to reconstruct the entire field of view (e.g.,
21\arcsec\ at $\lambda\,=\,1$~mm or about the size of the interior of an
internetwork region) simultaneously. Rather, the effective resolution may be 
limited by the number of pieces of information measured instantaneously: 2016
resolution elements (due to the number of baselines) correspond to an effective 
resolution of about 0\farcs5 at $\lambda\,=\,1$ mm.

A major advantage of ALMA for solar research lies in the properties of its 
wavelength range because (i) gas temperature linearly translates into emergent 
continuum intensity and (ii) synthetic intensity images for comparison can be 
calculated comparatively easily.

%================================================================================
\section{The hydrodynamic model}

Here we use the numerical 3D model described in Paper~I which has been calculated 
with the radiation hydrodynamics code \mbox{CO$^5$BOLD} (\cite{freytag02}).
Magnetic  fields are not included, restricting the model to internetwork 
regions. The resolution of the model atmosphere is 40~km in horizontal ($x$, $y$) 
and 12~km in vertical direction ($z$). The top of the model is located at a 
height of $z = 1710$~km in the middle chromosphere.  
The origin of the $z$~axis is defined by the level where the average Rosseland 
optical depth reaches unity. 
Since the horizontal extension is 5600~km, the model represents a quadratic patch 
of $7.\hspace*{-0.7mm}\arcsec7 \times 7.\hspace*{-0.7mm}\arcsec7$ of a 
non-magnetic internetwork region like it should be observable with ALMA. 

The model chromosphere is dominated by shock waves which are excited in the lower 
layers. Propagation and interaction of these waves give the model chromosphere 
its characteristic appearance: thin filaments of hot gas (shock waves) and 
embedded cool regions (see Fig.~\ref{fig.intimages}a). The mesh size of this 
network-like pattern is comparable to spatial scales typical for granulation.  
The average chromospheric temperature is roughly 3700~K but spans a large range 
from only $\sim 2000$~K to over 7000~K in shock waves. 
Caused by the large velocity of these waves, the whole pattern changes on time 
scales of only 20-25~s ($1/e$ autocorrelation time scale). 

As mentioned in Paper~I the radiative transfer in the simulation still needs to 
be improved for chromospheric conditions whereas the lower layers are modelled 
very realistically (see, e.g., Leenaarts \& Wedemeyer-B\"ohm, submitted).
The absolute temperature amplitudes are thus somewhat uncertain and need to be  
calibrated against observations. In contrast, the general topology of the upper 
layers (up to heights where magnetic fields become non-negligible) is a robust 
feature. The shock waves, which produce the pattern, are generated in the lower 
well-modelled layers, and their propagation and interaction does not sensitively 
depend on the thermodynamics of the chromospheric layers.   
Although the temperature amplitudes must be considered with caution, the model 
nevertheless provides insight in the involved spatial and temporal scales
and thus can help to define constraints for future observations.

%-------------------------------------------------------------------------------- 
\begin{figure}[t] 
\centering 
  \resizebox{\hsize}{!}{\includegraphics{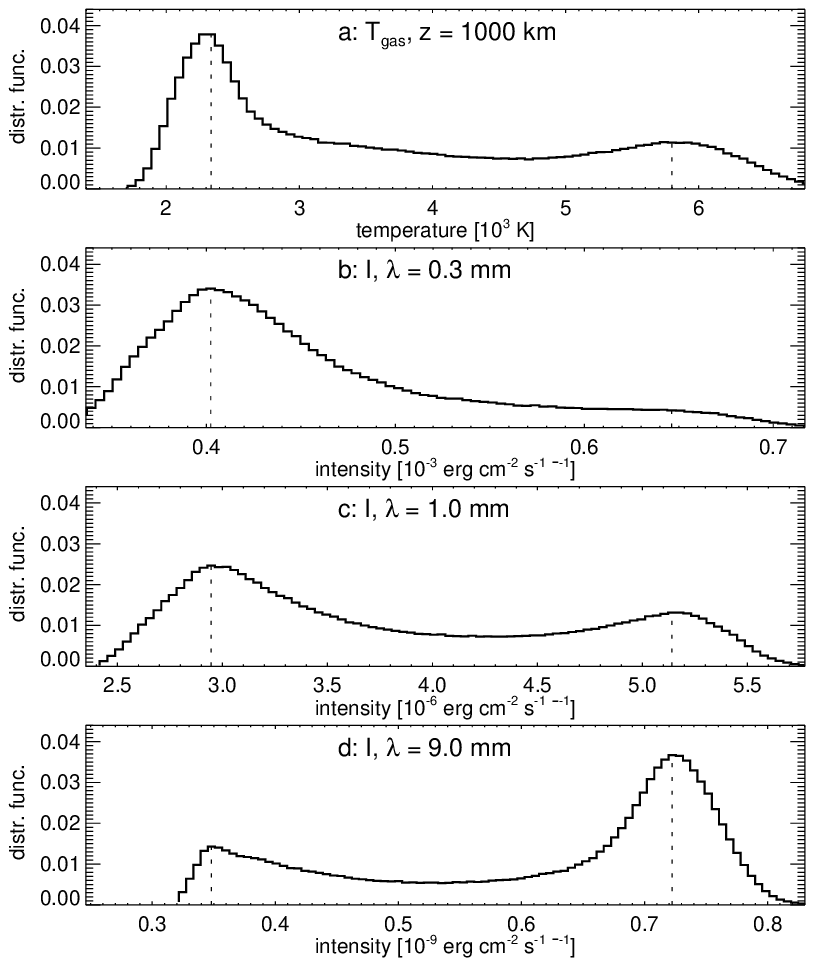}}
  \caption{
    Histograms for \textbf{a)} gas temperature in a horizontal slice at 
    geometrical height $z = 1000$~km, and \textbf{b-d)} emergent continuum 
    intensity at wavelengths of $0.3$~mm, $1$~mm, and 
    $9$~mm, respectively.} 
  \label{fig.inthist}
\end{figure} 

\begin{figure}[t] 
\centering 
  \resizebox{\hsize}{!}{\includegraphics{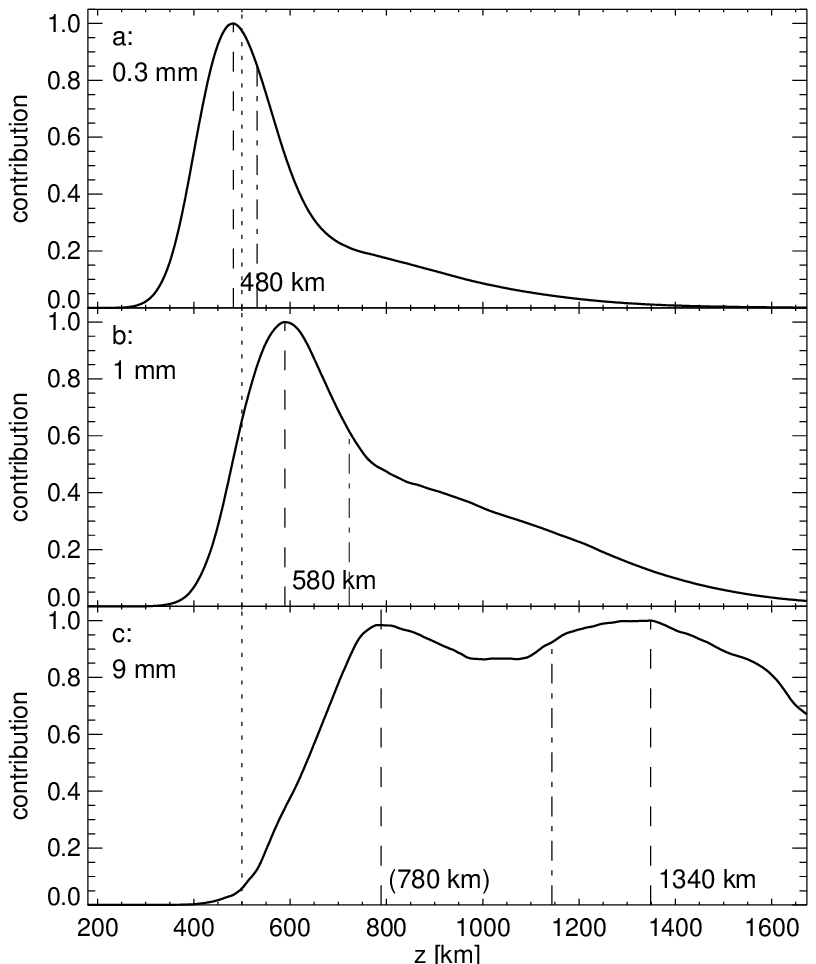}}
  \caption{Normalised averaged intensity contribution functions on the 
   geometrical height scale of the simulation sequence 
   at wavelengths of 
   \textbf{a)} $0.3$~mm, 
   \textbf{b)} $1$~mm, and
   \textbf{c)} $9$~mm. 
   The vertical lines mark the height of maximum contribution (dashed), 
   of half the integrated function (dot-dashed), and the 
   boundary photosphere/chromosphere (dotted).} 
  \label{fig.formheight}
\end{figure} 
%-------------------------------------------------------------------------------- 

%================================================================================
\section{Radiative transfer calculations}

The continuum at (sub-)millimeter wavelengths originates from the low and middle 
chromosphere and is mainly due to thermal free-free radiation. The 
assumption of local thermodynamic equilibrium (LTE) is valid and the source 
function is Planckian. Furthermore, $h\nu \ll k_\mathrm{B}T$, allowing to use 
the Rayleigh-Jeans approximation. Consequently, the source function is linearly 
related to the local gas temperature at these wavelengths. These conditions 
largely simplify the necessary calculations.  

The continuous opacity at mm and sub-mm wavelengths goes roughly proportional to 
the electron pressure. \cite*{carlsson02} showed that the ionisation fraction of 
hydrogen --- becoming the major electron contributor at chromospheric heights --- 
exhibits substantial deviations from thermodynamic equilibrium conditions. The 
ionisation fraction of hydrogen cannot follow the rapid changes of the 
thermodynamic conditions associated with the passage of shock waves but tends to 
stay on a fixed level characteristic for the hot, shocked gas phase. In 
comparison to LTE conditions this leads to a systematic increase of the opacity, 
particularly in the cool, largely neutral regions. On average we expect a shift 
of the emitting layers towards larger heights, and an increase of intensity
contrasts. The important deviations from thermodynamic equilibrium of the 
hydrogen ionisation have not been taken into account in the spectrum synthesis 
calculations of this preliminary study but should be considered in more detailed 
calculations. We nevertheless think that the relative spatial and temporal 
statistics of the intensity fluctuations is still reasonably represented by our 
simplified model.

The continuum images, which are presented here (see Fig.~\ref{fig.intimages}), 
were computed with the 3D LTE spectrum synthesis code LINFOR3D, which was 
developed by Steffen \& Ludwig (based on the Kiel code LINFOR/LINLTE). 
For convenience we adopt the following wavelengths for the spectrum synthesis, 
covering the range accessible by ALMA: 
0.3~mm \mbox{($\sim$~1000~GHz)}, 1~mm \mbox{($\sim$~300~GHz)}, and 9~mm 
\mbox{($\sim$~33~GHz)}. 
All calculations are done for disk-center ($\mu = 1.0$).

%================================================================================
\section{Results}

The horizontal temperature cross-section in Fig.1a shows a pattern which is 
characteristic for the model chromosphere. The intensity maps in 
\mbox{Fig.~\ref{fig.intimages}b-d} look more or less similar but with different 
contributions of dark and bright areas as can also be seen in the histograms in 
Fig.~\ref{fig.inthist}. The gas temperature distribution at $z = 1000$~km 
(panel a) exhibits two separate peaks, one at high temperatures due to hot shock 
waves and one at low values, caused by the cool background. The model shows a 
similar picture for all chromospheric heights above $\sim 800$~km where shock 
waves become clearly visible (see Paper~I for more details). The layers 
below are characterised by a single peak with a height-dependent width. 
The differences in the intensity histograms are therefore directly caused 
by the mapping of different height ranges, from where the continuum intensity 
emerges, thus sampling different sets of gas temperatures.
 
The height ranges can be seen in Fig.~\ref{fig.formheight} in the form of 
intensity contribution functions. At 0.3~mm the maximum intensity contribution 
originates from the upper photosphere ($z = 480$~km), whereas the chromosphere 
contributes only little compared to the intensity at 1~mm. At that wavelength 
the middle chromosphere produces a significant fraction, while the maximum is 
still located in the very low chromosphere ($z = 580$~km). 
Consequently, the intensity distribution at 1~mm is much more similar to the 
gas temperature at $z = 1000$~km compared to $\lambda = 0.3$~mm which mostly 
samples the high photosphere.

At 9~mm, however, the intensity is due to a large height range 
throughout the whole model chromosphere, even exceeding its vertical extent. 
Hence, a significant part of the emergent radiation at $\lambda = 9$~mm would 
come from layers not included in the model, so that the results derived for 
this wavelength should be interpreted with great care. 

The clear tendency of sampling higher layers as one goes to longer
wavelength is expected since the opacity scales as the wavelength
squared. The rapid increase of the opacity towards longer wavelength
is associated with a shift of the continuum forming layers towards
correspondingly lower density.

In principle, the temporal evolution of the chromospheric structure depends on 
the sampled height range, too. But since the evolution time scale (see Paper~I) 
stays roughly constant throughout the model chromosphere, the intensity maps 
evolve on the same time scales (20-25~s). 

Finally, we degraded the spatial resolution with a Gauss profile. 
The small-scale pattern is clearly visible at  $0.\hspace*{-0.7mm}\arcsec5$  
(FWHM). At $1\arcsec$, however, the image is very blurred but still 
exhibits fluctuations on granular scales. But the intensity contrast decreases  
strongly with resolution, complicating the detection of the pattern at 
resolutions above $1\arcsec$. 
Pushing the effective resolution as high as possible is thus clearly of crucial 
importance. These matters are discussed in more detail in a forthcoming 
paper.

%================================================================================ 
\section{Conclusions}

A comparison with future ALMA observations will form a strong test for present 
and future models of the solar atmosphere. 
As shown with these exemplary calculations, emergent continuum intensity at 
different wavelengths samples different height ranges of the (model) chromosphere. 
Hence, simultaneous observations at multiple wavelengths will provide substanial 
new insight in the hitherto hotly debated structure of the solar chromosphere. 
Given this powerful tool, also the propagation of shock waves and chromospheric 
oscillation modes can be addressed. 
However, as we did show in this work, high spatial and temporal resolution 
are of crucial importance for observing the small-scale structure of 
internetwork regions, whereas the field of view just needs to be large enough to 
cover a reasonable part of an internetwork region.

%================================================================================ 
\begin{acknowledgements}
We thank T.~Ayres, R.~Osten, and S.~White for helpful comments. 
\end{acknowledgements}

%================================================================================ 

%================================================================================ 

\begin{thebibliography}{}

\bibitem[\protect\astroncite{Asensio Ramos et~al.}{2003}]{asensio03}
  {Asensio Ramos}, A., {Trujillo Bueno}, J., {Carlsson}, M., \& {Cernicharo}, J.,
  2003, ApJL, 588, L61

\bibitem[\protect\astroncite{Ayres}{2002}]{ayres02}
  {Ayres}, T. 2002, ApJ, 575, 1104

\bibitem[\protect\astroncite{Bastian}{2002}]{bastian02}
  {Bastian}, T.~S. 2002, Astron.Nachr., 323, 271

\bibitem[\protect\astroncite{Carlsson \& Stein}{1995}]{carlsson95}
  {Carlsson}, M. \& {Stein}, R.~F. 1995, ApJL, 440, L29

\bibitem[\protect\astroncite{Carlsson \& Stein}{2002}]{carlsson02}
{Carlsson}, M. \& {Stein}, R.~F. 2002, ApJ, 572, 626

\bibitem[\protect\astroncite{Freytag et~al.}{2002}]{freytag02}
{Freytag}, B., {Steffen}, M., \& {Dorch}, B. 2002, Astron.~Nachr., 323, 213

\bibitem[\protect\astroncite{Fontenla et~al.}{1993}]{fal93}
  {Fontenla}, J.~M., {Avrett}, E.~H., \& {Loeser}, R. 1993, ApJ, 406, 319

\bibitem[\protect\astroncite{Leenaarts \&  Wedemeyer-B\"ohm}{submitted}]{leenaarts04}
  {Leenaarts}, J., \& {Wedemeyer-B\"ohm}, S., A\&A, submitted

\bibitem[\protect\astroncite{Loukitcheva et~al.}{2004}]{loukitcheva04}
  {{Loukitcheva}, M., {Solanki}, S.~K., {Carlsson}, M., \& {Stein}, R.~F.}, 2004, A\&A 419, 747 

\bibitem[\protect\astroncite{Uitenbroek}{2000}]{uitenbroek00a}
{Uitenbroek}, H. 2000, ApJ, 531, 571

\bibitem[\protect\astroncite{Vernazza et~al.}{1981}]{val81}
  {Vernazza}, J.~E., {Avrett}, E.~H., \& {Loeser}, R. 1981, ApJS, 45, 635

\bibitem[\protect\astroncite{Wedemeyer et~al.}{2004}]{wedemeyer04}
  {Wedemeyer}, S., {Freytag}, B., {Steffen}, M., {Ludwig}, H.-G., \& {Holweger}, H. 2004, A\&A 414, 1121

\end{thebibliography}
\end{document}